\documentclass[prd,amsfonts,noshowpacs,nofootinbib]{revtex4}
\usepackage{amsmath}

\newcommand{\formula}[2]{\begin{equation}\label{#1} #2 \end{equation}}

\begin{document}

\begin{abstract}
We consider a 5D BPS dilatonic two brane model which reduces to the
Randall-Sundrum model or the Ho\v{r}ava-Witten theory for a
particular choice of parameters. Recently new dynamical solutions
were found by Chen \emph{et al.}, which describe a moduli
instability of the warped geometry. Using a 4D effective theory
derived by solving the 5D equations of motion, based on the gradient
expansion method, we show that the exact solution of Chen \emph{et
al.} can be reproduced within the 4D effective theory and we
identify the origin of the moduli instability. We revisit the
gradient expansion method with a new metric ansatz to clarify why
the 4D effective theory solution can be lifted back to an exact 5D
solution. Finally we argue against a recent claim that the 4D
effective theory allows a much wider class of solutions than the 5D
theory and provide a way to lift solutions in the 4D effective
theory to 5D solutions perturbatively in terms of small velocities
of the branes.

\end{abstract}

\title{Moduli instability in warped compactification - 4D effective theory approach}
\author{Frederico Arroja\footnote{Frederico.Arroja@port.ac.uk}}
\author{Kazuya Koyama\footnote{Kazuya.Koyama@port.ac.uk}}
\affiliation{Institute of Cosmology and Gravitation, University of
Portsmouth, Portsmouth PO1 2EG, UK}
\date{\today}
\maketitle

\section{\label{sec:INTRO}Introduction}
The idea of using the degrees of freedom of extra spatial dimensions
to unify gravity and the other interactions has attracted much
interest. Especially, M-Theory has the potential to achieve this
ultimate goal. Ho\v{r}ava and Witten showed that M-Theory
compactified as $S^1/Z_2 \times M^{10}$ reduces to a known string
theory ($E_8\times E_8$ Heterotic String) \cite{HEE}. From a low
energy effective theory point of view, this model has a 5D spacetime
(the bulk) with two 4D boundary hypersurfaces (the branes)
\cite{LUK}. The remaining six spatial dimensions are assumed to be
compactified. All the Standard Model particles are confined to the
branes while gravity can propagate in the bulk. As a consequence of
the compactification of the six spatial dimensions, a 5D effective
scalar field appears in the bulk, which describes the volume of the
compactified 6D space.

A novel aspect of this model is that the fifth dimension is not
homogeneous. The line element for a static solution is given in the
form $ds_5^2=h^{\alpha}(r)dr^2+h^{\beta}(r)ds_4^2(x)$, where $r$ is
the fifth coordinate, $\alpha$, $\beta$ are constants and $h$ is
called the warp factor. This warped geometry is used by Randall and
Sundrum \cite{JUN} to address the hierarchy problem. They have also
showed that it is possible to localize gravity around the brane with
the warped geometry, providing in this way an alternative to
compactification \cite{MAY} (see Ref. \cite{REV} for a review).

Recently, a new class of dynamical solutions that describes an
instability of the warped geometry has been found. Chen \emph{et
al.} noticed that it is possible to obtain a dynamical solution by
replacing the constant modulus in the warp factor $h$ by a linear
function of the 4D coordinates \cite{ESM}. This solution describes
an instability of the model as the brane will hit the singularity in
the bulk, where $h=0$. This kind of solution exists also in the 10D
type IIB supergravity \cite{HK1}.

In this paper, we study the moduli instability in a dilatonic two
brane model, where the potentials for the scalar field on the brane
and in the bulk obey the BPS condition \cite{CVE}. For particular
values of parameters we retrieve either the Ho\v{r}ava-Witten theory
or the Randall-Sundrum model. We identify the origin of the moduli
instability using a 4D effective theory derived in Ref. \cite{KK}
(see Refs. \cite{moduli} and \cite{K} for different approaches).
This effective theory is derived by solving the 5D equations of
motion using the gradient expansion method \cite{gradient}, where we
assume that the velocities of the branes are small compared with the
curvature scale of the bulk determined by the warp factor.

Despite the fact that the 4D effective theory is based on a
slow-motion approximation, we will show that the 5D exact solutions
can be reproduced in the 4D effective theory. In order to understand
the relation between 4D solutions in an effective theory and full 5D
solutions, we revisit the gradient expansion method by employing a
new metric ansatz. Using this metric ansatz, we can clearly see why
the moduli instability solution in the 4D effective theory can be
lifted to an exact 5D solution.

We also comment on a recent claim that the 4D effective theory
allows a much wider class of solutions than the 5D theory
\cite{HK2}. We disagree with that conclusion and we show that it is
based on the restricted form of the 5D metric ansatz used in Ref.
\cite{HK2}. Using our metric ansatz, we provide a way to lift
solutions in the 4D effective theory to 5D solutions perturbatively
in terms of small velocities of the branes.

The structure of the paper is as follows. In section II, the model
used in this paper is described in detail. In section III, we
identify the solution in the 4D effective theory that describes the
moduli instability. In section IV, we revisit the gradient expansion
method to derive the 4D effective theory. We propose a new metric
ansatz which is useful to relate 4D solutions in the effective
theory to 5D solutions. Using this formalism we explain why the 4D
solution for the moduli instability can be lifted to an exact 5D
solution. In section V, we comment on the arguments against the 4D
effective theory. Section VI is devoted to conclusions.

\section{\label{sec:MODEL}The model}

Our model consists of a 5D spacetime (bulk) filled with a scalar
field. The fifth dimension is a compact space $S^1$ with a $Z_2$
symmetry (i.e. identification $r\rightarrow -r$) and will be
parameterized by the coordinate $r$, $-L\leq r\leq L$. The bulk
action takes the form
\formula{Sbulk}{S_{bulk}=\frac{1}{2\kappa_G^2}\int
d^5x\sqrt{-g_5}\left[{}^5\!R-\nabla_M\varphi\nabla^M\varphi+V_{bulk}(\varphi)\right],}
where $\kappa_G$ denotes the 5D gravitational constant. Throughout
this work latin indices can have values in $\{r,t,x,y,z\}$, while
greek indices do not include the extra dimension coordinate $r$. The
scalar field potential will be
\formula{Vbulk}{V_{bulk}(\varphi)=-\left(b^2-\frac{2}{3}\right)e^{-2\sqrt{2}b\varphi}\sigma^2,}
where $\sigma$ and $b$ are the remaining parameters of this model.
We can retrieve either the Randall-Sundrum model ($b=0$) or the
Ho\v{r}ava-Witten model ($b=1$) according to the value of $b$. The
orbifold $S^1/Z_2$ has two fixed points, at $r=0$ and $r=L$, and we
can put branes there. The branes' action is
\formula{Sbranes}{S_{branes}=S_++S_-,} where $S_+$ and $S_-$ are the
positive and negative tension brane action, respectively. They are
given by \formula{Sbrane+-}{S_\pm=\mp\frac{\sqrt{2}}{\kappa_G^2}\int
d^4x\sqrt{-g_4}e^{-\sqrt{2}b\varphi} \sigma.} The action of our
model is \formula{ACTT}{S_{total}=S_{bulk}+S_{branes}.}

\section{\label{sec:GEM}Exact solutions for moduli instability
in 4D effective theory}

\subsection{\label{subsec:SOL}The exact 5D solution}

Exact static solutions for the equations of motion that result from
(\ref{ACTT}) were obtained in \cite{CVE}. Recently, an interesting
dynamical solution was found by Chen \emph{et al.} \cite{ESM}. They
noticed that if we add a linear function of time to the warp factor
of the solution of \cite{CVE}, it would still be a solution of the
equations of motion. Their exact 5D solution reads
\begin{eqnarray}
dS_5^2&=&\left(h\tau-\frac{|r|}{l}\right)^\frac{6b^2-2}{3b^2+1}dr^2+\left(h\tau-\frac{|r|}{l}\right)^{\frac{2}{3b^2+1}}\eta_{\mu\nu}dx^\mu
dx^\nu,\label{ESMSM}
\\
\varphi&=&\frac{3\sqrt{2}b}{3b^2+1}\ln
\left(h\tau-\frac{|r|}{l}\right), \quad l=\frac{3}{3b^2+1}\frac{\sqrt{2}}{\sigma}, \label{ESMSD}
\end{eqnarray}
where $h$ is an arbitrary constant.

From the above equations we can read the scale factor of the
positive tension brane ($r=0$) as
\formula{sf+}{a^2_+(\tau)=\left(h\tau\right)^\frac{2}{3b^2+1},} and
the scalar field as
\formula{scf+}{\varphi(r=0,\tau)=\frac{3\sqrt{2}b}{3b^2+1}\ln(h\tau).}

Let us choose $h<0$ and $\tau <0$. In this case the proper distance
between branes, the so-called radion, is decreasing according to
\formula{RAD}{\mathcal{R}=\int_0^L\left(h\tau-\frac{r}{l}\right)^{\frac{3b^2-1}{3b^2+1}}dr=l\frac{3b^2+1}{6b^2}(h\tau)^\frac{6b^2}{3b^2+1}\left[1-\left(1-\frac{L/l}{h\tau}\right)^{\frac{6b^2}{3b^2+1}}\right].}
When $\tau\rightarrow-\infty$, $\mathcal{R}\rightarrow
L\left(h\tau\right)^\frac{3b^2-1}{3b^2+1}$ and for $\tau=L/hl$,
$\mathcal{R}=l(3b^2+1)\left(L/l\right)^\frac{6b^2}{3b^2+1}/6b^2$.
Before the two branes collide, a curvature singularity will appear
in the negative tension brane ($r=L$) at $\tau=L/hl$. This
singularity will move towards the positive tension brane and will
reach it at $\tau=0$. This event represents the total annihilation
of the spacetime.

It is useful to note that if we drop the modulus sign in
(\ref{ESMSM}), the bulk spacetime is a static black brane or black
hole solution depending on $b$, see \cite{ESM} for more details. If
$b=1$, there is a timelike curvature singularity at $h\tau=|r|/l$.
From the static bulk point of view, the two branes are moving in
this static bulk and the negative tension brane first hits the
singularity. Even if the bulk spacetime is static, the existence of
the branes, which gives the modulus sign for $r$, makes the
spacetime truly time dependent.

Moduli instability is a serious problem for these types of model. In
this work, we will try to understand it from a 4D effective theory
viewpoint.
\subsection{\label{subsec:ACT}The exact solutions in 4D effective theory}

In \cite{KK}, Kobayashi and Koyama applied the gradient expansion
method to solve perturbatively  the 5D equations of motion resulting
from (\ref{ACTT}). Their $0^{th}$ order solution reads
\begin{eqnarray}
dS_5^2&=&d(t)^2e^{2\sqrt{2}b\phi(t)}dr^2+\mathcal{F}(r,t)^\frac{1}{3b^2}
h_{\mu\nu}(t)dx^\mu dx^\nu,\label{KK0th1}
\\
\varphi(r,t)&=&\frac{1}{\sqrt{2}b}\ln \mathcal{F}(r,t)+\phi(t),
\quad \mathcal{F}(r,t)=1-\sqrt{2}b^2d(t)\sigma|r|.
\label{KK0th2}
\end{eqnarray}
The main result of the paper is a set of equations involving only
four dimensional quantities that describe the dynamics of the
unknown functions $d(t)$, $h_{\mu\nu}(t)$ and $\phi(t)$, appearing
in Eqs. (\ref{KK0th1}) and (\ref{KK0th2}). It can be shown that
their dynamical equations can be consistently deduced from the
following action \formula{EFFACT}{S_{eff}=\frac{l}{2\kappa_G^2}\int
d^4x\sqrt{-h}e^{\sqrt{2}b\phi}\left[\psi
R(h)-\frac{3}{2(1+3b^2)}\frac{1}{1-\psi}\nabla_\alpha\psi\nabla^\alpha\psi-
\psi\nabla_\alpha\phi\nabla^\alpha\phi\right],} where here
$\nabla_\alpha$ means the covariant derivative with respect to
$h_{\mu\nu}$ and $\psi=1-\left(1-d\right)^\frac{3b^2+1}{3b^2}$. This
is the action we obtain if we substitute Eqs. (\ref{KK0th1}) and
(\ref{KK0th2}) into (\ref{ACTT}) and integrate over the extra
dimension \cite{moduli}.

Performing the conformal transformation
$h_{\mu\nu}=e^{-\sqrt{2}b\phi}f_{\mu\nu}/\psi$ and defining new
scalar fields as
\formula{psi1}{\psi=1-\tanh^2\left[\sqrt{\frac{1+3b^2}{6}}\Psi\right],}
\formula{phih}{\phi=\frac{\Theta}{\sqrt{1+3b^2}}+\lambda\ln\psi,
\quad \lambda\equiv-\frac{3}{\sqrt{2}}\frac{b}{3b^2+1},} we arrive
at the action in the Einstein frame
\formula{EFFACTEF}{S_E=\frac{l}{2\kappa_G^2}\int
d^4x\sqrt{-f}\left[R(f)
-\Theta_{|\alpha}\Theta^{|\alpha}-\Psi_{|\alpha}\Psi^{|\alpha}\right].}

It is clear that the moduli fields have no potential and this is the
origin of the instability. We are interested in cosmological
solutions so we choose $f_{\mu\nu}$ to be a flat FRW metric.  The
Einstein equations resulting from the action (\ref{EFFACTEF}) are
\formula{Gxxtt}{\frac{a''}{a}=-\frac{1}{6}\left({\Psi'}^2+{\Theta'}^2\right),
\quad
\left(\frac{a'}{a}\right)^2=\frac{1}{6}\left({\Psi'}^2+{\Theta'}^2\right).}
and the equations of motion for the scalar fields are
\formula{PsieqThetaeq}{\Psi''+2\frac{a'}{a}\Psi'=0,\quad
\Theta''+2\frac{a'}{a}\Theta'=0,} where $a(\tau)$ is the FRW scale
factor and the prime denotes derivative with respect to the
conformal time $\tau$. Equations (\ref{Gxxtt},\ref{PsieqThetaeq})
can be easily integrated to find \formula{a2}{ \Psi(\tau)=
\sqrt{\frac{6}{3b^2+1}} \alpha \ln a^2+ \gamma,\quad \Theta(\tau)=
\sqrt{1+3 b^2} \beta \ln a^2+ \delta, \quad a^2(\tau)=\xi \tau+
\zeta,} \formula{zzzz}{ \alpha^2 + \frac{(1+3b^2)^2}{6} \beta^2 =
1+3 b^2,} where $\alpha$, $\beta$, $\gamma$, $\delta$, $\zeta$ and
$\xi$ are integration constants. The solutions for the fields in the
original frame can be obtained as
\formula{d1}{d(\tau)=1-\left[\frac{(\mu \tau+\nu)^{2
\alpha}-1}{(\mu\tau+\nu)^{2 \alpha}+1}\right]^\frac{6b^2}{3b^2+1},}
\formula{phi1}{\phi(\tau)= \ln\left[\frac{(\mu\tau+\nu)^{2 \alpha
\lambda(b) +\beta}} {\left[(\mu \tau +\nu)^{2
\alpha}+1\right]^{2\lambda(b)}} \right]+ \varrho,} where
$\lambda(b)$ is defined in Eq.~(\ref{phih}) and integration
constants $\mu$, $\nu$, $\varrho$ are redefined from $\xi$, $\zeta$,
$\gamma$ and $\delta$. The radion is calculated as
\formula{radionKK}{\mathcal{R}(\tau)=\frac{1}{\sqrt{2}b^2\sigma}d(\tau)e^{\sqrt{2}b\phi(\tau)},}
and the square of the scale factor on the positive tension brane is
\formula{sf+KK}{a^2_+(\tau)=\frac{e^{-\sqrt{2}b\phi(\tau)}}{\psi(\tau)}a^2(\tau).}
We have found the remarkable fact that for a particular choice of
the integration constants we can reproduce the exact solution found
by Chen \emph{et al.} described in the previous section. If we
choose integration constants obeying the relations
\formula{relations}{\alpha=\frac{1}{2}, \;\; \beta=-\lambda(b), \;\;
\mu= \frac{2 h l}{L}, \;\; \nu=-1,} we get the same 4D quantities
(scale factor and scalar field on the positive tension brane) and
radion as the Chen \emph{et al.} solution, Eqs.~(\ref{sf+},
\ref{scf+}, \ref{RAD}).

For this solution, the scale factor in the Einstein frame is given by
\begin{equation}
a^2(\tau)=\frac{2L}{l}\left(h \tau - \frac{L}{2l} \right).
\end{equation}
If we take $h<0$, $\tau <0$, this corresponds to a collapsing
universe, due to the kinetic energy of the scalar fields. At $\tau =
L/2hl <0$, the universe in the Einstein frame reaches the Big-bang
singularity. However, we should be careful to interpret this
singularity. In fact, in the original 5D theory, $\tau = L/2hl$ does
not correspond to any kind of singularity. In 5D theory, the
negative tension brane hits the singularity at $\tau = L/hl$ and the
positive tension brane hits the singularity at $\tau=0$. In fact at
$\tau = L/2hl$, $\psi =0$ and so the conformal transformation
becomes singular and the Einstein frame metric loses physical
meaning.

\section{\label{sec:GEMN}Gradient expansion method with a new metric ansatz}
In the previous section, we found that the exact solution derived in
\cite{ESM} can be reproduced from the $0^{th}$ order of the
perturbative method. In this section, we will revisit the gradient
expansion method which is used to derive the effective theory to
clarify the reason why the exact solution derived in \cite{ESM} can
be reproduced within the 4D effective theory.

\subsection{\label{subsec:5DEQ}5D Equations}
In this subsection we study the 5D equations of motion. From the
action (\ref{ACTT}) we derive the Einstein's equations
\begin{eqnarray}
{}^5G_{AB}&=&\nabla_A\varphi\nabla_B\varphi+\frac{1}{2}g_{AB}\left[-\nabla^C\varphi\nabla_C\varphi+V_{bulk}(\varphi)\right]
\nonumber \\
&&+\sqrt{2}\sigma\left[-\frac{\sqrt{g_4}}{\sqrt{g_5}}g_{\mu\nu}\delta_A^\mu\delta_B^\nu
e^{-\sqrt{2}b\varphi}\delta(r)+\frac{\sqrt{g_4}}{\sqrt{g_5}}g_{\mu\nu}\delta_A^\mu\delta_B^\nu
e^{-\sqrt{2}b\varphi}\delta(r-L)\right]. \label{Einstein1}
\end{eqnarray}

Assuming that the 5D line element has the form
\formula{metric1}{dS_5^2=g_{AB}(X)dX^AdX^B=g_{rr}(x,r)dr^2+g_{\mu\nu}(x,r)dx^\mu
dx^\nu,} ($x$ dependence means dependence of all the other
coordinates $\{x,y,z,t\}$, except the extra dimension $r$) we can
extract the junction conditions for the metric tensor from
Einstein's equation,
\formula{JC}{\sqrt{g^{rr}}\left(K_\nu^\mu-\delta_\nu^\mu
K\right){\!}\Bigg|_{\stackrel{\scriptstyle{r=0^+}}{\scriptstyle{r=L^-}}}=\mp\frac{1}{\sqrt{2}}\sigma
e^{-\sqrt{2}b\varphi}\delta_\nu^\mu\Bigg|_{\stackrel{\scriptstyle{r=0^+}}{\scriptstyle{r=L^-}}},}
where the extrinsic curvature $K_{\mu\nu}$ is defined by
$K_{\mu\nu}=-\frac{1}{2}\partial_rg_{\mu\nu}$. The scalar field
equation of motion is
\formula{scalarfeq}{\Box\varphi+\sqrt{2}b\left(b^2-\frac{2}{3}\right)\sigma^2e^{-2\sqrt{2}b\varphi}=-2b\sigma\left[\sqrt{g^{rr}}e^{-\sqrt{2}b\varphi}\delta(r)-\sqrt{g^{rr}}e^{-\sqrt{2}b\varphi}\delta(r-L)\right],}
and so the junction conditions for the scalar field are
\formula{JCSF}{\sqrt{g^{rr}}\partial_r\varphi{\!}\Bigg|_{\stackrel{\scriptstyle{r=0^+}}{\scriptstyle{r=L^-}}}=\mp
b\sigma
e^{-\sqrt{2}b\varphi}\Bigg|_{\stackrel{\scriptstyle{r=0^+}}{\scriptstyle{r=L^-}}}.}

In order to proceed we shall assume that the 5D line element has
further symmetries, described by
\formula{metric2}{dS_5^2=e^{2\sqrt{2}b\tilde{\varphi}(x,r)}H^{\frac{6b^2-2}{3b^2+1}}(x,r)dr^2+H^{\frac{2}{3b^2+1}}(x,r)\tilde{g}_{\mu\nu}dx^\mu
dx^\nu,\quad H(x,r)=C(x)-\frac{r}{l}.} In this section, we will
assume that the position of the second brane is $r=L=l$.  For the
scalar field we will assume the form
\formula{sfd}{\varphi(x,r)=\frac{3\sqrt{2}b}{3b^2+1}\ln
H+\tilde{\varphi}(x,r).} This metric ansatz is inspired by the time
dependent solution (\ref{ESMSM},\ref{ESMSD}) of Chen \emph{et al.}
\cite{ESM}. Their solution was found by replacing the modulus
parameter in the static solution by a linear function of time. In
the same manner, we introduce an $x$ dependence in $H$ through the
modulus parameter $C(x)$ in a covariant way . We also introduce the
function $\tilde{\varphi}$ for the scalar field moduli. In order to
satisfy the junction conditions (\ref{JCSF}) we must have the
exponential factor in the $g_{rr}$ metric component. The tensor
$\tilde{g}_{\mu\nu}$ is left completely general.

After some mathematical manipulations of the Einstein equations
we obtain
\begin{eqnarray}
&&\frac{1}{l}e^{-2\sqrt{2}b\tilde{\varphi}}H^{\frac{-9b^2+1}{3b^2+1}}\left(\frac{3b^2-5}{3b^2+1}\tilde{K}^\mu_\nu-\frac{1}{3b^2+1}\delta^\mu_\nu\tilde{K}\right)
+e^{-2\sqrt{2}b\tilde{\varphi}}H^{\frac{-6b^2+2}{3b^2+1}}\left(\tilde{K}^\mu_{\nu,r}-\tilde{K}\tilde{K}^\mu_\nu\right)
\nonumber \\
&&-\sqrt{2}be^{-2\sqrt{2}b\tilde{\varphi}}H^{\frac{-3b^2+1}{3b^2+1}}\tilde{\varphi}_{,r}\left(\frac{1/l}{3b^2+1}H^{\frac{-6b^2}{3b^2+1}}\delta^\mu_\nu+H^{\frac{-3b^2+1}{3b^2+1}}\tilde{K}^\mu_\nu\right)
-H^{\frac{-3b^2-3}{3b^2+1}}\left(C_{|\nu}^{|\mu}+\frac{1}{3b^2+1}C_{|\alpha}^{|\alpha}\delta_\nu^\mu\right)
\nonumber \\
&&-H^{-\frac{2}{3b^2+1}}\left[\sqrt{2}b\tilde{\varphi}_{|\nu}^{|\mu}+\left(2b^2+1\right)\tilde{\varphi}^{|\mu}\tilde{\varphi}_{|\nu}\right]
-\frac{\sqrt{2}b}{3b^2+1}H^{\frac{-3b^2-3}{3b^2+1}}\left[\left(3b^2+1\right)\left(\tilde{\varphi}^{|\mu}C_{|\nu}+\tilde{\varphi}_{|\nu}C^{|\mu}\right)+\tilde{\varphi}^{|\alpha}C_{|\alpha}\delta_\nu^\mu\right]
\nonumber \\
&&+H^{-\frac{2}{3b^2+1}}R_\nu^\mu(\tilde{g})=0
,\label{munueq1}
\end{eqnarray}

\begin{eqnarray}
&&\frac{1}{l}\frac{3b^2-3}{3b^2+1}e^{-2\sqrt{2}b\tilde{\varphi}}H^{\frac{-9b^2+1}{3b^2+1}}\tilde{K}
+e^{-2\sqrt{2}b\tilde{\varphi}}H^{\frac{-6b^2+2}{3b^2+1}}\left(\tilde{K}_{,r}-\tilde{K}^{\alpha\beta}\tilde{K}_{\alpha\beta}\right)
\nonumber \\
&&+e^{-2\sqrt{2}b\tilde{\varphi}}H^{\frac{-6b^2+2}{3b^2+1}}\tilde{\varphi}_{,r}\left(\frac{1}{l}\frac{2\sqrt{2}b}{3b^2+1}H^{-1}-\sqrt{2}b\tilde{K}-\tilde{\varphi}_{,r}\right)
\nonumber \\
&&-H^{\frac{-3b^2-3}{3b^2+1}}\left(\frac{3b^2-1}{3b^2+1}C_{|\alpha}^{|\alpha}+\frac{6\sqrt{2}b^3}{3b^2+1}\tilde{\varphi}^{|\alpha}C_{|\alpha}\right)
-\sqrt{2}bH^{-\frac{2}{3b^2+1}}\left(\tilde{\varphi}_{|\alpha}^{|\alpha}+\sqrt{2}b\tilde{\varphi}^{|\alpha}\tilde{\varphi}_{|\alpha}\right)=0
,\label{rreq1}
\end{eqnarray}

\begin{eqnarray}
&&\tilde{K}^{\alpha}_{\beta|\alpha}-\tilde{K}_{|\beta}+\sqrt{2}b\left(\tilde{\varphi}_{|\beta}\tilde{K}-\tilde{\varphi}_{|\alpha}\tilde{K}^{\alpha}_{\beta}\right)
+H^{-1}\left(\frac{3b^2-2}{3b^2+1}C_{|\beta}\tilde{K}-\frac{3b^2-5}{3b^2+1}C_{|\alpha}\tilde{K}^{\alpha}_{\beta}\right)
\nonumber \\
&&=-\tilde{\varphi}_{,r}\left(\frac{3\sqrt{2}b}{3b^2+1}H^{-1}C_{|\beta}+\tilde{\varphi}_{|\beta}\right)
,\label{rbetaeq1}
\end{eqnarray}
where $\tilde{K}_{\mu\nu}$ is defined like
$\tilde{K}_{\mu\nu}=-\frac{1}{2}\partial_r{\tilde{g}_{\mu\nu}}$,
$\tilde{K}$ is the trace of $\tilde{K}_{\mu\nu}$,
$\scriptstyle{|}_{\scriptstyle{\alpha}}$ denotes covariant
derivative with respect to $\tilde{g}_{\mu\nu}$ and
$R_{\mu\nu}(\tilde{g})$ is the Ricci tensor of $\tilde{g}_{\mu\nu}$.
Equation (\ref{scalarfeq}) transforms into
\begin{eqnarray}
&&e^{-2\sqrt{2}b\tilde{\varphi}}H^{\frac{-3b^2+5}{3b^2+1}}\left[\tilde{\varphi}_{,rr}-\sqrt{2}b\left(\tilde{\varphi}_{,r}\right)^2-\tilde{\varphi}_{,r}\tilde{K}+\frac{1}{l}\frac{1}{3b^2+1}H^{-1}\left(3\sqrt{2}b\tilde{K}+\left(9b^2-5\right)\tilde{\varphi}_{,r}\right)\right]
\nonumber \\
&&+H\tilde{\varphi}_{|\alpha}^{|\alpha}+\frac{3\sqrt{2}b}{3b^2+1}C_{|\alpha}^{|\alpha}+\sqrt{2}bH\tilde{\varphi}_{|\alpha}\tilde{\varphi}^{|\alpha}
+\frac{9b^2+1}{3b^2+1}\tilde{\varphi}_{|\alpha}C^{|\alpha}=0
.\label{sfeq1}
\end{eqnarray}
With this particular ansatz the junction conditions
(\ref{JC},\ref{JCSF}) are significantly simplified and read
\formula{jcKuv}{\left[\tilde{K}_\nu^\mu\right]{\!}\Bigg|_{\stackrel{\scriptstyle{r=0^+}}{\scriptstyle{r=l^-}}}=0,}
\formula{jcsf}{\left[\tilde{\varphi}_{,r}\right]{\!}\Bigg|_{\stackrel{\scriptstyle{r=0^+}}{\scriptstyle{r=l^-}}}=0.}

It is impossible to solve these equations in general, so
in the next section, we will solve them using the gradient expansion
method up to first order in the perturbations.
\subsection{\label{subsec:PM}The gradient expansion}
\subsubsection{\label{subsubsec:APPR}The approximation}

In order to use perturbation theory to solve a system of
differential equations we need to identify the characteristic
scale of the different terms involved in the equations and then
see if there is a small parameter.

The derivatives along the extra dimension of the conformal metric
$\tilde{g}_{\mu\nu}$ as well as the derivative of $\tilde{\varphi}$
are of order $1/l$. We assume that variations along the branes'
coordinates are small in comparison with $1/l$. This implies that
the radion changes slowly or that the speed of the branes is small.
More precisely, our small parameters will be
\formula{approx}{l^2\partial^2_x(\ldots)\ll1 \;\;\;\;\;\mbox{and}
\;\;\;\;\;\left(l\partial_x\ldots\right)^2\ll1,} where $\ldots$
represents the conformal metric functions, $C$ or $\tilde{\varphi}$.

As in the usual perturbation method, we expand the unknown functions
in a series
\formula{expanm}{\tilde{g}_{\mu\nu}(r,x)=\stackrel{(0)}{\tilde{g}}_{\mu\nu}(r,x)+
\stackrel{(1)}{\tilde{g}}_{\mu\nu}(r,x)+\cdots,}
\formula{expansf}{\tilde{\varphi}(r,x)=\stackrel{(0)}{\tilde{\varphi}}(r,x)+
\stackrel{(1)}{\tilde{\varphi}}(r,x)+\cdots.} We impose the boundary
conditions at the position of the positive tension brane
\formula{bcm}{\stackrel{(n)}{\tilde{g}}_{\mu\nu}(r=0,x)=0,\;\mbox{for
all}\;n>0,}
\formula{bcsf}{\stackrel{(n)}{\tilde{\varphi}}(r=0,x)=0,\;\mbox{for
all}\;n>0.} Other quantities are naturally expanded as
\formula{oexpan}{\tilde{K}_{\mu\nu}=\stackrel{(0)}{\tilde{K}}_{\mu\nu}+\stackrel{(1)}{\tilde{K}}_{\mu\nu}+\cdots.}

\subsubsection{\label{subsubsec:BACK}$0^{th}$ Order (Background geometry)}

The $0^{th}$ order system can be easily integrated with respect to
the extra dimension coordinate $r$ to get the particular solution
\formula{0sol}{\stackrel{(0)}{\tilde{g}_{\mu\nu}}(r,x)=\stackrel{(0)}{\tilde{g}_{\mu\nu}}(x),}
\formula{0solsf}{\stackrel{(0)}{\tilde{\varphi}}(r,x)=\stackrel{(0)}{\tilde{\varphi}}(x).}
This solution clearly satisfies the $0^{th}$order junction
conditions.

\subsubsection{\label{subsubsec:1ST}$1^{st}$ Order}

At $1^{st}$ order the evolution equations are
\begin{eqnarray}
  &&\frac{1}{l}e^{-2\sqrt{2}b\stackrel{(0)}{\tilde{\varphi}}}H^{\frac{-9b^2+1}{3b^2+1}}\left(\frac{3b^2-5}{3b^2+1}\stackrel{(1)}{\tilde{K}^\mu_\nu}-\frac{1}{3b^2+1}\delta^\mu_\nu\stackrel{(1)}{\tilde{K}}\right)
+e^{-2\sqrt{2}b\stackrel{(0)}{\tilde{\varphi}}}H^{\frac{-6b^2+2}{3b^2+1}}\stackrel{(1)}{\tilde{K}^\mu_{\nu,r}}
-\frac{\sqrt{2}b}{l}\frac{1}{3b^2+1}e^{-2\sqrt{2}b\stackrel{(0)}{\tilde{\varphi}}}H^{\frac{-9b^2+1}{3b^2+1}}\stackrel{(1)}{\tilde{\varphi}_{,r}}\delta^\mu_\nu
  \nonumber \\
  &&-H^{\frac{-3b^2-3}{3b^2+1}}\left(C_{|\nu}^{|\mu}+\frac{1}{3b^2+1}C_{|\alpha}^{|\alpha}\delta_\nu^\mu\right)
-H^{-\frac{2}{3b^2+1}}\left[\sqrt{2}b\stackrel{(0)}{\tilde{\varphi}_{|\nu}^{|\mu}}+\left(2b^2+1\right)\stackrel{(0)}{\tilde{\varphi}^{|\mu}}\stackrel{(0)}{\tilde{\varphi}_{|\nu}}\right]
  \nonumber \\
  &&-\frac{\sqrt{2}b}{3b^2+1}H^{\frac{-3b^2-3}{3b^2+1}}\left[\left(3b^2+1\right)\left(\stackrel{(0)}{\tilde{\varphi}^{|\mu}}C_{|\nu}+\stackrel{(0)}{\tilde{\varphi}_{|\nu}}C^{|\mu}\right)+\stackrel{(0)}{\tilde{\varphi}^{|\alpha}}C_{|\alpha}\delta_\nu^\mu\right]
+H^{-\frac{2}{3b^2+1}}R_\nu^\mu\left(\stackrel{(0)}{\tilde{g}}\right)=0
,\label{1storder1}
\end{eqnarray}
\begin{eqnarray}
  &&\frac{1}{l}\frac{3b^2-3}{3b^2+1}e^{-2\sqrt{2}b\stackrel{(0)}{\tilde{\varphi}}}H^{\frac{-9b^2+1}{3b^2+1}}\stackrel{(1)}{\tilde{K}}
+e^{-2\sqrt{2}b\stackrel{(0)}{\tilde{\varphi}}}H^{\frac{-6b^2+2}{3b^2+1}}\stackrel{(1)}{\tilde{K}_{,r}}
+\frac{1}{l}\frac{2\sqrt{2}b}{3b^2+1}e^{-2\sqrt{2}b\stackrel{(0)}{\tilde{\varphi}}}H^{\frac{-9b^2+1}{3b^2+1}}\stackrel{(1)}{\tilde{\varphi}_{,r}}
  \nonumber \\
  &&-H^{\frac{-3b^2-3}{3b^2+1}}\left(\frac{3b^2-1}{3b^2+1}C_{|\alpha}^{|\alpha}+\frac{6\sqrt{2}b^3}{3b^2+1}\stackrel{(1)}{\tilde{\varphi}^{|\alpha}}C_{|\alpha}\right)
-\sqrt{2}bH^{-\frac{2}{3b^2+1}}\left(\stackrel{(0)}{\tilde{\varphi}_{|\alpha}^{|\alpha}}+\sqrt{2}b\stackrel{(0)}{\tilde{\varphi}^{|\alpha}}\stackrel{(0)}{\tilde{\varphi}_{|\alpha}}\right)=0
,\label{1storder2}
\end{eqnarray}
\begin{eqnarray}
  &&e^{-2\sqrt{2}b\stackrel{(0)}{\tilde{\varphi}}}H^{\frac{-3b^2+5}{3b^2+1}}\left[\stackrel{(1)}{\tilde{\varphi}_{,rr}}+\frac{1}{l}\frac{1}{3b^2+1}H^{-1}\left(3\sqrt{2}b\stackrel{(1)}{\tilde{K}}+\left(9b^2-5\right)\stackrel{(1)}{\tilde{\varphi}_{,r}}\right)\right]
\nonumber \\
&&+H\stackrel{(0)}{\tilde{\varphi}_{|\alpha}^{|\alpha}}+\frac{3\sqrt{2}b}{3b^2+1}C_{|\alpha}^{|\alpha}+\sqrt{2}bH\stackrel{(0)}{\tilde{\varphi}_{|\alpha}}\stackrel{(0)}{\tilde{\varphi}^{|\alpha}}
+\frac{9b^2+1}{3b^2+1}\stackrel{(0)}{\tilde{\varphi}_{|\alpha}}C^{|\alpha}=0
.\label{1storder3}
 \end{eqnarray}
The junction conditions at this order are
\formula{jc1}{\left[\stackrel{(1)}{\tilde{K}_\nu^\mu}\right]{\!}\Bigg|_{\stackrel{\scriptstyle{r=0^+}}{\scriptstyle{r=l^-}}}=0,}
\formula{jc1sf}{\left[\stackrel{(1)}{\tilde{\varphi}_{,r}}\right]{\!}\Bigg|_{\stackrel{\scriptstyle{r=0^+}}{\scriptstyle{r=l^-}}}=0.}
In the preceding equations all the indices are raised with the
zeroth order metric. Combining the trace of equation
(\ref{1storder1}) with  equation (\ref{1storder2}) we obtain
\formula{K1}{\frac{1}{l}e^{-2\sqrt{2}b\stackrel{(0)}{\tilde{\varphi}}}\stackrel{(1)}{\tilde{K}}
=-\left(C_{|\alpha}^{|\alpha}+\sqrt{2}bC_{|\alpha}\stackrel{(0)}{\tilde{\varphi}^{|\alpha}}\right)H^{\frac{6b^2-4}{3b^2+1}}
+\frac{3b^2+1}{6}\left[R\left(\stackrel{(0)}{\tilde{g}}\right)-\stackrel{(0)}{\tilde{\varphi}_{|\alpha}}\stackrel{(0)}{\tilde{\varphi}^{|\alpha}}\right]H^{\frac{9b^2-3}{3b^2+1}}
-\frac{\sqrt{2}b}{l}e^{-2\sqrt{2}b\stackrel{(0)}{\tilde{\varphi}}}\stackrel{(1)}{\tilde{\varphi}_{,r}}.}
Imposing the junction conditions (\ref{jc1},\ref{jc1sf}) we get
\formula{riccis}{\tilde{R}\left(\stackrel{(0)}{\tilde{g}}\right)=\stackrel{(0)}{\tilde{\varphi}_{|\alpha}}\stackrel{(0)}{\tilde{\varphi}^{|\alpha}},}
and the equation of motion for the 4D effective scalar field
\formula{aa}{C_{|\alpha}^{|\alpha}+\sqrt{2}bC_{|\alpha}\stackrel{(0)}{\tilde{\varphi}^{|\alpha}}=0.}
Equation (\ref{K1}) now reads
\formula{K10}{\stackrel{(1)}{\tilde{K}}=-\sqrt{2}b\stackrel{(1)}{\tilde{\varphi}_{,r}}.}

Using the decomposition of $\stackrel{(1)}{\tilde{K_\nu^\mu}}$ in
\formula{deco}{\stackrel{(1)}{\tilde{K_\nu^\mu}}=\stackrel{(1)}{\tilde{\Sigma_\nu^\mu}}
+\frac{1}{4}\delta_\nu^\mu\stackrel{(1)}{\tilde{K}},} equation
(\ref{1storder1}) can be easily integrated to find
\begin{equation}
 \begin{array}{l}
e^{-2\sqrt{2}b\stackrel{(0)}{\tilde{\varphi}}}H^{-\frac{3b^2-5}{3b^2+1}}\stackrel{(1)}{\tilde{\Sigma_\nu^\mu}}=
\\
\left[\left(C_{|\nu}^{|\mu}+\sqrt{2}b\left(C_{|\nu}\stackrel{(0)}{\tilde{\varphi}^{|\mu}}+C^{|\mu}\stackrel{(0)}{\tilde{\varphi}_{|\nu}}\right)\right)r
-\left(R_\nu^\mu\left(\stackrel{(0)}{\tilde{g}}\right)-\sqrt{2}b\stackrel{(0)}{\tilde{\varphi}_{|\nu}^{|\mu}}-\left(2b^2+1\right)\stackrel{(0)}{\tilde{\varphi}_{|\nu}}\stackrel{(0)}{\tilde{\varphi}^{|\mu}}\right)\left(Cr-\frac{r^2}{2l}\right)\right]_{traceless}
+\chi_\nu^\mu(x) ,
 \end{array}\label{sigma1}
\end{equation}
where $\chi_\nu^\mu(x)$ is an integration constant and the subscript
$[\;\;]_{traceless}$ means the traceless part of the quantity between
square brackets. In terms of $\tilde{\Sigma_\nu^\mu}$, the junction
conditions (\ref{jc1}) are
\formula{jcsigma1}{\left[\stackrel{(1)}{\tilde{\Sigma}_\nu^\mu}\right]{\!}\Bigg|_{\stackrel{\scriptstyle{r=0^+}}{\scriptstyle{r=l^-}}}=0.}

From the previous junction conditions (\ref{jcsigma1}) we can obtain
the 4D effective equations of motion
\begin{eqnarray}
\tilde{R}_\nu^\mu\left(\stackrel{(0)}{\tilde{g}}\right)&=&\left(C-\frac{1}{2}\right)^{-1}\left[C_{|\nu}^{|\mu}+\frac{1}{4}\delta_\nu^\mu
C_{|\alpha}^{|\alpha}+\sqrt{2}b\left(C_{|\nu}\stackrel{(0)}{\tilde{\varphi}^{|\mu}}+C_{|\mu}\stackrel{(0)}{\tilde{\varphi}^{|\nu}}\right)\right]
+\sqrt{2}b\left(\stackrel{(0)}{\tilde{\varphi}_{|\nu}^{|\mu}}-\frac{1}{4}\delta_\nu^\mu\stackrel{(0)}{\tilde{\varphi}_{|\alpha}^{|\alpha}}\right)
\nonumber\\
&&+\left(2b^2+1\right)\stackrel{(0)}{\tilde{\varphi}_{|\nu}}\stackrel{(0)}{\tilde{\varphi}^{|\mu}}-\frac{b^2}{2}\delta_\nu^\mu\stackrel{(0)}{\tilde{\varphi}_{|\alpha}}\stackrel{(0)}{\tilde{\varphi}^{|\alpha}}
,\label{4deffectiveeq}
\end{eqnarray}
and \formula{chi}{\chi_\nu^\mu(x)=0.}

Combining equation (\ref{K10}) with the scalar field equation
(\ref{1storder3}) and integrating it once with respect to the extra
dimension, we get (after using the previous equations of motion to
simplify the result)
\formula{drsf}{e^{-2\sqrt{2}b\stackrel{(0)}{\tilde{\varphi}}}H^{\frac{-3b^2+5}{3b^2+1}}\stackrel{(1)}{\tilde{\varphi}_{,r}}=-\left(Cr-\frac{r^2}{2l}\right)\left(\stackrel{(0)}{\tilde{\varphi}_{|\alpha}^{|\alpha}}+\sqrt{2}b\stackrel{(0)}{\tilde{\varphi}_{|\alpha}}\stackrel{(0)}{\tilde{\varphi}^{|\alpha}}\right)-\stackrel{(0)}{\tilde{\varphi}^{|\alpha}}C_{|\alpha}r+\Xi(x),}
where $\Xi(x)$ is just an integration constant.

The junction conditions (\ref{jc1sf}) give \formula{Xi}{\Xi(x)=0}
and the equation of motion for the second 4D effective scalar field
\formula{2eq}{\stackrel{(0)}{\tilde{\varphi}_{|\alpha}^{|\alpha}}+\sqrt{2}b\stackrel{(0)}{\tilde{\varphi}_{|\alpha}}\stackrel{(0)}{\tilde{\varphi}^{|\alpha}}=-\left(C-\frac{1}{2}\right)^{-1}\stackrel{(0)}{\tilde{\varphi}^{|\alpha}}C_{|\alpha}.}

\subsection{\label{subsec:ACT2}The 4D Effective theory}
The 4D effective equations of motion are summarized as
\begin{eqnarray}
\tilde{R}_\nu^\mu\left(\stackrel{(0)}{\tilde{g}}\right)&=&\left(C-\frac{1}{2}\right)^{-1}\left[C_{|\nu}^{|\mu}+\frac{1}{4}\delta_\nu^\mu
C_{|\alpha}^{|\alpha}+\sqrt{2}b\left(C_{|\nu}\stackrel{(0)}{\tilde{\varphi}^{|\mu}}+C_{|\mu}\stackrel{(0)}{\tilde{\varphi}^{|\nu}}\right)\right]
+\sqrt{2}b\left(\stackrel{(0)}{\tilde{\varphi}_{|\nu}^{|\mu}}-\frac{1}{4}\delta_\nu^\mu\stackrel{(0)}{\tilde{\varphi}_{|\alpha}^{|\alpha}}\right)
\nonumber\\
&&+\left(2b^2+1\right)\stackrel{(0)}{\tilde{\varphi}_{|\nu}}\stackrel{(0)}{\tilde{\varphi}^{|\mu}}-\frac{b^2}{2}\delta_\nu^\mu\stackrel{(0)}{\tilde{\varphi}_{|\alpha}}\stackrel{(0)}{\tilde{\varphi}^{|\alpha}}
,\label{4deffectiveeqA}
\end{eqnarray}
\formula{sfR}{C_{|\alpha}^{|\alpha}+\sqrt{2}bC_{|\alpha}\stackrel{(0)}{\tilde{\varphi}^{|\alpha}}=0,}
\formula{2eqC}{\stackrel{(0)}{\tilde{\varphi}_{|\alpha}^{|\alpha}}+\sqrt{2}b\stackrel{(0)}{\tilde{\varphi}_{|\alpha}}\stackrel{(0)}{\tilde{\varphi}^{|\alpha}}=-\left(C-\frac{1}{2}\right)^{-1}\stackrel{(0)}{\tilde{\varphi}^{|\alpha}}C_{|\alpha},}
and they can be deduced from the following action
\formula{action2}{S_{eff}=\frac{l}{\kappa_G^2}\int
d^4x\sqrt{-\stackrel{(0)}{\tilde{g}}}\left(C-\frac{1}{2}\right)
e^{\sqrt{2}b\stackrel{(0)}{\tilde{\varphi}}}\left[R(\stackrel{(0)}{\tilde{g}})-\stackrel{(0)}{\tilde{\varphi}_{|\alpha}}\stackrel{(0)}{\tilde{\varphi}^{|\alpha}}\right],}
where $\scriptstyle{|}_{\scriptstyle{\alpha}}$ denotes covariant
derivative with respect to $\stackrel{(0)}{\tilde{g}_{\mu\nu}}$. We
should note that this effective action can be derived by
substituting in (\ref{ACTT}) the 5D solutions up to the first order
and integrating it over the fifth dimension \cite{K}.

As a consistency check, we see that if we perform the conformal
transformation
\formula{1}{h_{\mu\nu}(x)=C^{\frac{2}{3b^2+1}}(x)\stackrel{(0)}{\tilde{g}_{\mu\nu}}(x),}
the previous action reduces to (\ref{EFFACT}) if the effective
scalar fields of the two theories are related through
\formula{2}{\psi=2C^{-2}\left(C-\frac{1}{2}\right),}
\formula{3}{\phi(x)=\frac{3\sqrt{2}b}{3b^2+1}\ln C
+\stackrel{(0)}{\tilde{\varphi}}(x).} The check consists in seeing
that these relations are exactly the ones required so that the two
observables (the proper distance between branes and the scalar field
on the positive tension brane) agree in both approaches.

\subsection{\label{subsec:SOL2}The 5D exact solution}

It is straightforward to find a cosmological solution of this 4D
effective theory. For example we can easily find the following particular
solution
\begin{equation}
\stackrel{(0)}{\tilde{g}_{\mu\nu}}(x)=\eta_{\mu\nu},\quad
C(x)=ht,\quad \stackrel{(0)}{\tilde{\varphi}}(x)=0,
\label{4dsolution}
\end{equation}
where $h$ is an integration constant.

Now we are ready to address the question why the above solution can
be lifted to an exact 5D solution. Let us start by calculating the
next order correction ${}^{(1)}{\tilde{g}_{\mu\nu}}$ and
${}^{(1)}{\tilde{\varphi}}$. Eq. (\ref{drsf}) and the boundary
conditions (\ref{bcsf}) give ${}^{(1)}{\tilde{\varphi}}=0$, if we
take as $0^{th}$ order solution Eqs. (\ref{4dsolution}). We can
construct ${}^{(1)}{\tilde{K}_{\mu\nu}}$ from Eqs. (\ref{K10}) and
(\ref{sigma1}). For the $0^{th}$ order solution (\ref{4dsolution})
this gives ${}^{(1)}{\tilde{K}_{\mu\nu}}=0$. After imposing the
boundary conditions (\ref{bcm}), we obtain that the next order
correction vanishes, ${}^{(1)}{\tilde{g}_{\mu\nu}(r,x)}=0$. For
solution (\ref{4dsolution}) it turns out that all the corrections
vanish and the $0^{th}$ order solution is an exact solution of the
non-perturbed 5D Eqs. (\ref{munueq1}-\ref{jcsf}).

For other solutions of the 4D effective theory, higher order
corrections will not vanish and therefore they should be taken into
account in the reconstruction of the 5D metric. Using the gradient
expansion method, we can reconstruct the 5D solution perturbatively.
We should emphasize that the choice of the $0^{th}$ order metric is
quite important in order to reconstruct 5D solutions efficiently.
Our metric ansatz has the advantage that it is possible to recover
the exact solution of Chen \emph{et al.} (\ref{4dsolution}) at
$0^{th}$ order. Indeed, if we had started with an ansatz like Eqs.
(\ref{KK0th1},\ref{KK0th2}) we would need an infinite number of
higher order terms to obtain the exact 5D solution.

\section{\label{sec:COMM}Validity of 4D effective theory}
In this section we will make comments on a recent work by Kodama and
Uzawa \cite{HK2}. Let us start by briefly describing their
arguments. After deriving the 4D effective theory for warped
compactification of the 5D Ho\v{r}ava-Witten model (they also extend
their analysis to 10D IIB supergravity and obtain the same
conclusions), the authors show that the 4D effective theory allows a
wider class of solutions than the fundamental higher dimensional
theory. Therefore we should be careful in using this effective
theory approach, because we may find 4D solutions that do not
satisfy the equations of motion once lifted back to 5D.

The authors assume a metric ansatz of the form
\formula{met}{dS_5^2=h(x,r)dr^2+h^{\frac{1}{2}}(x,r)\tilde{g}_{\mu\nu}(x)dx^\mu
dx^\nu,} where the warp factor has the form $h(x,r)=C(x)-r/L$. This
corresponds to taking $\tilde{K}_{\mu \nu}=0$, $\tilde{\varphi}=0$
and $b=1$ (for the Ho\v{r}ava-Witten case) in our work. Then Eq.
(\ref{munueq1}) reduces to
\begin{equation}
-H(r,x)^{-\frac{3}{2}}\left(C(x)_{|\nu}^{|\mu}+\frac{1}{4}C(x)_{|\alpha}^{|\alpha}
\delta_\nu^\mu\right)
+H(r,x)^{-\frac{1}{2}}R_\nu^\mu(\tilde{g}(x))=0.
\end{equation}
In order to satisfy this equation for all values of $r$, we should
have \formula{con}{R_{\mu\nu}(\tilde{g})=0,\quad C_{|\mu \nu}=0.}

They obtain the 4D effective action, by integrating the fifth
dimension, as \formula{actionHK}{S_{eff} \propto \int
d^4x\sqrt{-\tilde{g}}\left(C(x)-\frac{1}{2}\right) R(\tilde{g}),}
which agrees with our effective action (\ref{action2}). As we have
shown, this theory admits solutions with $R_{\mu\nu}(\tilde{g})\neq
0$ (see Eq.~(\ref{4deffectiveeqA})), which do not obey the constraint (\ref{con}) obtained from the
5D equations of motion.

However, it is clear from our analysis that their metric ansatz is
too restrictive.  If we consider a more general metric as our metric
ansatz, we see that the 5D Einstein equations contain more terms
given by $\tilde{K}_{\mu \nu}$. With the inclusion of these new
terms, the 5D equations do not necessarily imply (\ref{con}). Of
course, the non-vanishing $\tilde{K}_{\mu\nu}$ changes the metric
(\ref{met}) and one could argue that the resultant 4D effective
action would be also changed. However, it is shown that even if we
include the first order corrections
${}^{(1)}{\tilde{g}_{\mu\nu}}(x)$ to the metric, the resultant 4D
effective action derived by integrating out the fifth dimension does
not change \cite{K}. Therefore, for 4D solutions that do not satisfy
(\ref{con}), we should include the corrections to the metric
(\ref{met}). We have provided this correction perturbatively. Using
Eqs. (\ref{K1}) and (\ref{sigma1}), we can reconstruct the
correction to the metric, ${}^{(1)}{\tilde{g}_{\mu\nu}}(x)$, which
is necessary to satisfy the 5D equation of motion.

We should emphasize that the validity of the 4D effective theory is
based on the conditions (\ref{approx}). If the 4D effective theory
admits a solution that violates the conditions (\ref{approx}), then
there is no guarantee that the 4D solution can be lifted up to the
5D solution consistently. We should check the validity of the 4D
effective theory by calculating the higher order corrections to
ensure that the higher order corrections can be neglected
consistently.
\section{\label{sec:CON}Conclusion}
In this paper, we studied the moduli instability in a two brane
model with a bulk scalar field recently found by Chen \emph{et al.}.
This model can be viewed as a generalization of the
Ho\v{r}ava-Witten theory and the Randall-Sundrum model. The scalar
field potentials in the bulk and on the branes are tuned in order to
satisfy the BPS condition.

We used a low energy effective theory, which is derived by assuming
that variations along the brane coordinates of the metric are small
compared with variations along the dimension perpendicular to the
brane. The effective theory is a bi-scalar tensor theory where one
of the scalar fields arises from the bulk scalar field (dilaton) and
the other arises from the degree of freedom of the distance between
branes (radion). In the Einstein frame, the theory consists of two
massless scalar fields, and the lack of potentials for these moduli
fields was shown to be responsible for the instability.

We found that the exact solution derived in \cite{ESM} can be
reproduced from the $0^{th}$ order of the perturbative method,
despite the fact that slow-motion approximations are used. We
revisited the gradient expansion method which is used to derive the
effective theory, in order to understand why the exact solution
derived in \cite{ESM} can be reproduced within the 4D effective
theory. We proposed a new metric ansatz which is useful to see the
relation between the solutions in the effective theory and the full
solutions for 5D equation of motion. Using this metric ansatz, it is
transparent why the moduli instability solution can be lifted to a
full 5D solution. We have also shown that not all solutions in the
4D effective theory can be lifted to exact 5D solutions. For these
solutions, the solutions in the effective theory receive higher
order corrections in velocities of the branes and we need to find 5D
solutions perturbatively.

Finally, we comment on the recent arguments against the 4D effective
theory. Ref. \cite{HK2} claims that the 4D effective theory allows a
much wider class of solutions than the 5D theory. We argued that
this conclusion comes from a too restricted metric ansatz used in
Ref. \cite{HK2}. Using a more general metric ansatz, we provided a
way to reconstruct the full 5D solutions from the solutions in the
4D effective theory.

Our method can be applied to other warped compactifications such as
10D type IIB supergravity models. In fact, there have been debates
on the validity of the metric ansatz commonly used to derive the 4D
effective theory. Our 10D generalization of the $0^{th}$ order
metric ansatz agrees with that proposed in Ref. \cite{GM} and the
method presented in this paper will provide a consistent way to
reduce the 10D theory to the 4D effective theory based on this
metric ansatz. This will be reported in a future publication
\cite{future}.

\begin{acknowledgments}
We thank Roy Maartens for valuable suggestions which improved the
paper. The authors benefitted from discussions during the conference
``The Next Chapter in Einstein's Legacy" and the subsequent workshop
at the Yukawa institute, Kyoto, Japan in July 2005, where a part of
this work was presented by KK. FA is supported by ``Funda\c{c}\~{a}o
para a Ci\^{e}ncia e Tecnologia (Portugal)", with the fellowship's
reference number: SFRH/BD/18116/2004. KK is supported by PPARC.
\end{acknowledgments}


\end{document}